\documentclass{article}
\usepackage[top=1.5in, bottom=1.5in, left=1.2in, right=1.2in]{geometry}
\usepackage{authblk}
\usepackage{amssymb, amsmath,framed}
\usepackage{graphicx}
\usepackage[round]{natbib}

\usepackage{bm}
\usepackage[colorlinks=true,linkcolor=blue,urlcolor=blue,citecolor=blue,anchorcolor=blue]{hyperref}
\expandafter\ifx\csname package@font\endcsname\relax\else
 \expandafter\expandafter
 \expandafter\usepackage
 \expandafter\expandafter
 \expandafter{\csname package@font\endcsname}
\fi
\def\bq{\begin{equation}}
\def\eq{\end{equation}}
\def\bqy{\begin{eqnarray}}
\def\eqy{\end{eqnarray}}

\makeatother

\begin{document}

\title{Role of stellar physics in regulating the critical steps for life}

\author{Manasvi Lingam \thanks{Electronic address: \texttt{manasvi.lingam@cfa.harvard.edu}}}

\author{Abraham Loeb \thanks{Electronic address: \texttt{aloeb@cfa.harvard.edu}}}
\affil{Institute for Theory and Computation, Harvard University, 60 Garden St, Cambridge MA 02138, USA}

\date{}

\maketitle

\begin{abstract}
We use the critical step model to study the major transitions in evolution on Earth. We find that a total of five steps represents the most plausible estimate, in agreement with previous studies, and use the fossil record to identify the potential candidates. We apply the model to Earth-analogs around stars of different masses by incorporating the constraints on habitability set by stellar physics including the habitable zone lifetime, availability of ultraviolet radiation for prebiotic chemistry, and atmospheric escape. The critical step model suggests that the habitability of Earth-analogs around M-dwarfs is significantly suppressed. The total number of stars with planets containing detectable biosignatures of microbial life is expected to be highest for K-dwarfs. In contrast, we find that the corresponding value for intelligent life (technosignatures) should be highest for solar-mass stars. Thus, our work may assist in the identification of suitable targets in the search for biosignatures and technosignatures.
\end{abstract}

\section{Introduction} \label{SecIntro}
In less than a decade, our understanding of exoplanets has improved dramatically thanks to the \emph{Kepler} mission, which was launched in 2009 \citep{Bor10,Bat14,Bor16}. The fields of exoplanetary science and astrobiology also received two major boosts over the last couple of years as a result of two remarkable discoveries. The first was the discovery of the potentially habitable planet Proxima b around Proxima Centauri, the star nearest to our Solar system \citep{AAB16}. The second was the discovery of at least seven Earth-sized planets orbiting the star TRAPPIST-1 at a distance of about $12$ pc \citep{GJL16,GTD17}, some of which may be capable of hosting liquid water on their surfaces. Looking ahead, there are a wide range of space- and ground-based telescopes that will become operational within the next $10$-$15$ years with the express purpose of hunting for myriad exoplanetary biosignatures \citep{FAD,SK18}.

The Search for Extraterrestrial Intelligence (SETI) has also received an impetus in this period on both the theoretical and observational fronts \citep{Cab16}. Theoretically, many innovative technosignatures have been proposed for identifying artifacts of extraterrestrial species, both extant and extinct \citep{BCD11,WMS14,WCZ16}. From the observational standpoint, the recently established \emph{Breakthrough Listen} project \citep{WD17,IS17} has injected new funding and rejuvenated SETI,\footnote{\url{https://breakthroughinitiatives.org/initiative/1}} after the unfortunate demise of federal funding in 1993. 

Thus, from the perspective of searching for biosignatures and technosignatures, it is therefore necessary to understand what are the constraints imposed on planetary habitability by the host star. This will help in facilitating the optimal selection of suitable target stars and planets, where the prospects for life may be maximized \citep{HJ10,LL18,Mali18,KGO18}. In this paper, we will therefore use a model originally developed by \citet{Cart83}, where evolution is treated as a succession of critical steps, to assess the likelihood of primitive (microbial) and intelligent (technological) life, and the implications for detecting biosignatures and technosignatures. A brief description of the methodology is provided in Sec. \ref{SecMeth}, followed by an extended discussion of the critical step model in the context of Earth's evolutionary history in Sec. \ref{SecET}. Next, we assess the likelihood of these critical steps being successfully attained on other exoplanets in Sec. \ref{SSecExo}. We conclude with a summary of our major points in Sec. \ref{SecConc}.

\section{Methodology}\label{SecMeth}
We begin with a brief summary of the mathematical preliminaries. A detailed derivation of these results can be found in \citet{BT85}, \citet{Cart08} and \citet{Wat08}. In the critical-step model, the basic assumption is that there are $n$ critical (i.e. ``hard'') steps in all. Each step is stochastic in nature, and has an  associated probability of occurrence (denoted by $\lambda_i$ with $i = 1,\dots,n$), and the condition $\lambda_i t_H \ll 1$ must be satisfied $\forall\, i$. Here, $t_H$ denotes the total period of habitability, and its value for the Earth and other exoplanets will be addressed later.

The central quantity of interest is the probability density function (PDF) for the case where the $r$-th step takes place at time $t$, and the remaining $n-r$ steps take place after $t$. Denoting this quantity by $P_{r,n}(t)$, the PDF can be expressed as
\begin{equation}\label{PDF}
    P_{r,n}(t) = \frac{n !}{(n-r)! (r-1)!} \frac{t^{r-1}\left(t_H - t\right)^{n-r}}{t_H^n}.
\end{equation}
Hence, the mean time taken for the $r$-th step, represented by $\bar{t}_{r,n}$, is
\begin{equation}\label{MeanT}
    \bar{t}_{r,n} = \int_0^{t_H} t P_{r,n}(t)\,dt = \left(\frac{r}{n+1}\right)t_H,
\end{equation}
and hence it follows that the average spacing $\left(\Delta t_n\right)$ between two consecutive steps is approximately equal,\footnote{This important fact - along with the idea that this methodology could be used to assess the accuracy of models describing the major evolutionary transitions on Earth - was first emphasized by Robin Hanson in his unpublished manuscript on evolutionary transitions: \url{http://mason.gmu.edu/~rhanson/hardstep.pdf}} with
\begin{equation}\label{Space}
    \Delta t_n = \frac{t_H}{n+1}.
\end{equation}
The cumulative probability $\mathcal{P}_{r,n}(t)$ that the $r$-th step occurs at a time $\leq t$ is given by
\begin{equation}\label{CP}
    \mathcal{P}_{r,n}(t) = \frac{n !}{(n-r)! (r-1)!} B\left(t/t_H\,;\,r\,,\,n-r+1\right),
\end{equation}
where $B$ is the incomplete beta function. For the limiting case $r=n$, (\ref{CP}) reduces to $\mathcal{P}_{n,n}(t) = \left(t/t_H\right)^n$. 

\section{Critical Steps and Major Transitions on Earth}\label{SecET}
We briefly discuss the use of critical steps model as a heuristic for understanding the major breakthroughs in the evolutionary history of the Earth \citep{Lun13,Knoll15}.

Before proceeding further, we wish to emphasize that sequences of evolutionary transitions encountered henceforth that are not in strong agreement with the theoretical model presented in Sec. \ref{SecMeth} may nevertheless be ``correct'', since the real issue could stem from the mathematical framework employed herein (based on critical steps). In other words, the existence of alternative paradigms that do not envision evolution as a series of random critical steps remains a distinct possibility. Furthermore, it should be recognized that the classification of evolutionary steps into ``easy'' or ``hard'' (i.e. critical) categories is both abstract and binary in nature.

For example, in the ``long fuse'' model \citep{Bog11}, a series of likely steps (each with a non-negligible timescale) unfold, culminating in slow, but near-inevitable, evolution - as per this framework, the emergence of any particular evolutionary innovation essentially becomes a matter of time. Hence, in this particular scenario, low-mass stars would be ideally suited for the evolution of intelligent life because of their longer main-sequence lifetimes. A second scenario is the ``many paths'' model in which the probability of a given major evolutionary transition is enhanced due to the fact that exist a large number of trajectories that can culminate in the desired final outcome \citep{BaSM16}.

\subsection{How many critical steps were present?}\label{SSecNo}
Although this question has been explored recently by means of the critical steps approach \citep{Cart08,Wat08,ML10}, there are some major points of divergence in our analysis, as discussed below.

One notable difference is that we assume the Earth was habitable approximately $4.5$ Ga (Gyr ago), as opposed to previous treatments which specified the earliest point of habitability as $4$ Ga. The primary reason for the choice of $4$ Ga was motivated by the fact that the Late Heavy Bombardment (LHB) - a phase during which the Earth was subjected to cataclysmic bombardment by a high number of impactors \citep{GLT05} - was detrimental to habitability. However, there are several lines of evidence that now suggest that the LHB may not have been a significant impediment to habitability:
\begin{itemize}
    \item There is some evidence indicating that the cratering record may also be explained via a sustained declining bombardment, instead of the intense LHB \citep{BN17}. If this hypothesis is correct, the prospects for habitability are improved, and the Earth may have been geologically habitable as early as $\approx 4.5$ Ga \citep{VP02,ZA07,Har09,AN12}. For instance, if the bombardment was relatively moderate, it could even have served as a valuable energy source for prebiotic chemistry \citep{CS92,RBD14}, leading to the synthesis of biomolecules such as amino acids, peptides and nucleobases \citep{MP13,FNS15}.
    \item Even if the LHB were present, numerical models which computed the extent of crustal melting indicate that hyperthermophiles may have survived in near-surface and subsurface environments \citep{AM09,GM18}; see also \citet{SAL17}.
    \item Yet another possibility is that life-bearing ejecta spawned during the LHB can return to the Earth, and thereby reseed it over short ($\sim 10^3$ yr) timescales \citep{WAG03,GDL05}, effectively ensuring that habitability was almost continuously prevalent during the Hadean-Archean eons.
\end{itemize}

Thus, we start our habitability ``clock'' at $4.5$ Ga. Several studies have attempted to assess the end of Earth's habitability in the future due to the increasing solar luminosity and the onset of the greenhouse effect. While early models yielded a value of $0.5$ Gyr in the future \citep{LW82}, more recent analyses have pushed forward this boundary to $\approx 1$-$2$ Gyr in the future \citep{CK92,FBVB,LVB01,GW12}. While the $2$ Gyr limit is conceivably valid for extremophiles, the limits for more complex organisms (including humans) could be closer to $1$ Gyr \citep{FBV06,WT15}. In addition, there may be other astrophysical risks posed to habitability over multi-Gyr timescales \citep{BJ09,MT11,SAL17}. Hence, we will assume that the Earth becomes uninhabitable $\approx 1$ Gyr in the future, but we will address the $2$ Gyr case later in Sec. \ref{SSecExt}.

As per the preceding discussion, $t_H \approx 4.5 + 1 \approx 5.5$ Gyr. Let us suppose that the evolution of technological intelligence (i.e. \emph{Homo sapiens}) represents the $n$-th step, and use the fact that the mean timescale for our emergence was $\bar{t}_{n,n} \approx 4.5$ Gyr. From (\ref{MeanT}) and the above values, it follows that $n \approx 4.5$. Thus, it seems plausible that a 4- or 5-step model may represent the best fit. This result is in good agreement with earlier studies that arrived at the conclusion $n = 5$ \citep{Wat08,Cart08,ML10}, and we will adopt this value henceforth. Classical frameworks for understanding the course of evolution on Earth also seemingly indicate that the total number of critical steps was quite small \citep{Sch94,SS95,deDu,La09}, namely $n \lesssim 10$, and could have been $5$ to $6$ in number \citep{KB00,Jud17}.

\subsection{What were the five critical steps?}\label{SSecFive}
In order to determine the five critical steps, there are two routes that are open to us. The first approach assumes that these steps correspond to the major evolutionary transitions identified in the seminal work of \citet{JMS95,MS99}, wherein each step involves noteworthy changes in the storage and transmission of information. Such paradigms have been extensively invoked and utilized by several authors \citep{JL06,Koo7,CS11,WFG15,BaSM16,OMP16}. This strategy for identifying the critical steps was employed by \citet{Wat08}, who observed that the temporal constraints on the first three transitions (origin of replicating molecules, chromosomes, and the genetic code) indicate that not all of them are likely to be critical steps; instead, if the origin of prokaryotic cells is considered as a single critical step, the $n = 5$ model can be formulated accordingly.\\

\noindent{{\bf (1A) Origin of (Prokaryotic) Life:}} Of all the potential critical steps, dating the origin of life (abiogenesis) is the most difficult owing to the near-absence of sedimentary rocks and the action of processes like diagenesis and metamorphism \citep{KBS16}. We will adopt a conservative approach, and adopt the value of $3.7$ Ga for the earliest robust evidence of life. There are two independent lines of evidence that support this date. The first is the recent discovery of stromatolite-like structures in the Isua Supracrustal Belt (ISB) by \citet{NB16}. The second stems from the low $\delta^{13}$C values in graphite globules from the ISB \citep{Ros99,OK14}, which is conventionally indicative of biological activity. The oldest microfossils, which arguably display evidence of cell structure (e.g. lumen and walls), were discovered in the Pilbara Craton and date from $3.4$-$3.5$ Ga \citep{WK11,BAS15}. Here, it should be noted that even older claims for life do exist - the potentially biogenic carbon in a $4.1$ Ga Jack Hills zircon \citep{BBH15} and putative microfossils $> 3.8$ Ga in the Nuvvuagittuq belt \citep{DPG17} are two such examples - but they are not unambiguous. As per our discussion, the timescale for abiogenesis on Earth $\left(t_0\right)$ after the onset of habitability is $t_0 \approx 0.8$ Gyr. From (\ref{CP}), the cumulative probability is found to be $\mathcal{P}_{1A} = 0.54$.\\

\noindent{{\bf (2A) Origin of Eukaryotes:}} The origin of the crown eukaryotes is believed to have occurred through endosymbiosis \citep{Sag67,EM06,Ar15,LEM17} between an archaeon \citep{ESL17} - probably a member of the Lokiarchaeota \citep{SSJ15}, recently classified as belonging to the Asgard superphylum \citep{ZNC17} - and a proto-mitochondrion \citep{GBL99} that was closely related to $\alpha$-Proteobacteria \citep{PG14,PG16}. This event was apparently a very important one from the standpoint of bioenergetics and the eventual increase in biological complexity \citep{LM10,MGZ15}; see, however, \citet{BD15} and \citet{LyM15} for dissenting viewpoints. The oldest fossils that appear to be unambiguously eukaryotic in origin are the vesicles from the Changzhougou Formation and date from approximately $1.65$ Ga \citep{LAC09,LLS}. There are several other ostensibly eukaryotic microfossils that have been dated to $1.4$-$1.6$ Ga, and possibly as old as $1.8$ Ga \citep{HR92,Knoll14,DFB16,JK17,BSBW}. Phylogenetic molecular clock models have yielded ages for the Last Eukaryotic Common Ancestor (LECA) ranging between $1$ and $2$ Ga, although recent studies are closer to the latter value \citep{ES14,MOP14,LoMo15,SRPK17}. Although earlier claims for eukaryotic microfossils exist, for e.g. in the $2.1$ Ga Francevillian B Formation \citep{ABC10}, the $2.7$ Ga shales from the Pilbara Craton \citep{BLB99}, the Transvaal Supergroup sediments from $2.5$-$2.7$ Ga \citep{WSS09}, and the $2.7$-$2.8$ Ga lacustrine deposits of South Africa \citep{KK16},\footnote{It should be noted that some analyses based on molecular clocks have also concluded that eukaryogenesis took place $\gtrsim 2$ Ga \citep{HBVS,HK09,GCF17}.} we shall adopt the circumspect timing of $1.8$ Ga for the origin of eukaryotes. The corresponding timescale of $2.7$ Gyr leads us to the cumulative probability $\mathcal{P}_{2A} = 0.80$.\\

\noindent{{\bf (3A) Origin of Plastids:}} In the original list of major evolutionary transitions \citep{JMS95}, sexual reproduction was present in place of plastids. An important reason for this alteration was because there exists sufficiently compelling evidence that LECA was a complex organism that was capable of sexual reproduction \citep{Koo10,Butt15}; in other words, the origin of sexual reproduction was possibly coincident with eukaryogenesis \citep{Sza15,SLE15}, although there is no \emph{a priori} reason to believe that this apparent coincidence will always be valid on other inhabited exoplanets.\\
The importance of plastids stems from the fact that they enable eukaryotic photosynthesis. Eukaryotes acquired this ability by means of endosymbiosis with a cyanobacterium \citep{RBB05,Arch,Keel10}, thereby giving rise to the ``primary'' plastids in algae and plants \citep{GWM08,PCY12}. This endosymbiosis is believed to have occurred around $1.5$-$1.75$ Ga \citep{YHC,FKK04,RWB07,PLK11,OED14}, and these estimates appear to be consistent with the recent discovery of multicellular rhodophytes from $1.6$ Ga \citep{BSBW}. However, recent evidence based on molecular clock analyses favors the origin of the Archaeplastida (that possess plastids) by $1.9$ Ga \citep{SRPK17}. We choose to err on the side of caution and use $1.5$ Ga as the origin of the primary plastids. Upon calculating the cumulative probability using (\ref{CP}), we find $\mathcal{P}_{3A} = 0.58$.\\

\noindent{{\bf (4A) Origin of Complex Multicellularity:}} In this context, the rise of ``complex multicellularity'' refers to the emergence of plants, fungi and animals \citep{SS95}. An important point worth noting here is that each of these clades could have originated at a different time. The earliest evidence for metazoan fossils has been argued to be at least $0.64$ Ga \citep{LG09,MR10}, but it cannot be regarded as wholly conclusive. Molecular clocks indicate that the last common ancestor of animals lived around $0.8$ Ga or earlier \citep{DSB04,WLS,EF11,RK13,CLBD17}. The molecular clock evidence for plants suggests that their origins may extend as far back as $\approx 0.7$-$0.9$ Ga \citep{HGE01,LM04,CWD11,MHQ13}, although these methods are subject to much variability; the direct fossil evidence for plants is much more recent \citep{KN17}. Lastly, the use of molecular clocks to determine the origin of fungi has led to the estimate of $\approx 0.76$-$1.06$ Ga \citep{LHP09}. Thus, taken collectively, it seems plausible that the origin of complex multicellularity was about $0.8$ Ga \citep{Rok08}, although the discovery of \emph{Bangiomorpha pubescens}, whose age has been estimated to be $\lesssim 1.2$ Ga \citep{Butt00}, could be construed as evidence for an earlier divergence time. This hypothesis gains further credibility in light of the distinctive increase in eukaryotic diversity documented in the fossil record at $0.8$ Ga \citep{KJHC,Kno11}. The cumulative probability for this step is $\mathcal{P}_{4A} = 0.47$.\\

\noindent{{\bf (5A) Origin of Humans:}} More accurately, the revised version, \citet{Sza15} refers to the origin of ``Societies with natural language'', thus emphasizing the role of language. Since anatomically and behaviorally modern humans evolved only $\sim 10^5$ yr ago \citep{Kle95,Tat09,Ste11}, the timescale for the evolution of \emph{H. sapiens} (or even genus \emph{Homo}) since the onset of habitability is $4.5$ Gyr. Hence, the cumulative probability is estimated to be $\mathcal{P}_{5A} = 0.37$ by making use of (\ref{CP}).\\

Next, we shall outline the second strategy for deducing the five critical transitions. In order to do so, let us recall that the spacing between each critical step is roughly equal. From (\ref{Space}), we find that $\Delta t_n \approx 0.9$ Ga. Thus, if we can identify five ``important'' transitions, i.e. the breakthroughs that occurred only once or a handful of times, during Earth's geobiological and evolutionary history that have a spacing of $\approx 0.9$ Ga, they might potentially constitute the critical steps leading to technological intelligence. We will present our five transitions below, and offer reasons as to why they may happen to be the critical steps.\\

\noindent{{\bf (1B) Origin of Prokaryotic Life:}} Our choice of (1B) is the same as (1A). The issue of whether abiogenesis is an ``easy'' or a ``hard'' phenomenon remains currently unresolved \citep{IW17}, but resolving this question will have important implications for gauging the likelihood of extraterrestrial life \citep{LD02,Dav03,ST12,CK18}. However, in the spirit of most conventional analyses, we will suppose that abiogenesis does constitute one of the critical steps. In this case, the cumulative probability turns out to be $\mathcal{P}_{1B} = \mathcal{P}_{1A} = 0.54$.\\

\noindent{{\bf (2B) Origin of Oxygenic Photosynthesis:}} The evolution of oxygenic photosynthesis, due to the origin of prokaryotes akin to modern cyanobacteria \citep{MKM06}, had a profound impact on the Earth's biosphere \citep{HB11,FHV16}. On metabolic grounds, there are strong reasons to posit the emergence of oxygenic photosynthesis as a major transition in its own right \citep{LW11,OMP16}. The many advantages due to oxygenic photosynthesis have been succinctly summarized by \citet{Jud17}. The addition of oxygen to the atmosphere led to the formation of the ozone layer, caused an increase in the diversity of minerals, enabled the creation of new ecological niches, and, above all, aerobic metabolism releases about an order of magnitude more energy compared to anaerobic metabolism \citep{McC07,KB08}. The origin of oxygenic photoautotrophs remains very poorly constrained \citep{AM07} with chronologies ranging between $3.8$ Ga to $1.9$ Ga, with the former estimate arising from indirect evidence of environmental oxidation based on U–Th–Pb isotopic ratios \citep{RF04,Bui08,FCB16} and the latter representing the oldest direct evidence from microfossils \citep{FHJ16}. If we naively take the mean of these two values, we obtain $\approx 2.7$ Ga. There are several lines of evidence, although not all of them constitute robust biomarkers \citep{RFB08,FHH15,NNR16}, which appear to indicate that oxygenic photosynthesis evolved approximately $2.7$ Ga or later \citep{EF06,FMC10,FZB11,SCB12,PAH14,SGD15,SSW16,SHW17}, i.e. a few $100$ Myr prior to the onset of the Great Oxygenation Event (GOE). With the choice of $t = 1.8$ Gyr (which corresponds to $2.7$ Ga) for oxygenic photosynthesis, we obtain a cumulative probability of $\mathcal{P}_{2B} = 0.53$ after using (\ref{CP}).\\
As noted earlier, the origin of oxygenic photosynthesis has been subject to much controversy and uncertainty. Hence, it is quite conceivable that the GOE served as a critical step in the origin of complex (eukaryotic) life, and the attainment of sufficient oxygen levels could serve as an evolutionary bottleneck on exoplanets \citep{Kno85,CG05,mlal19}. The GOE was a highly significant event that led to a considerable enhancement of Earth's atmospheric oxygen levels to $\sim 1\%$ of the present-day value around $2.4$ to $2.1$ Ga \citep{LO16,GCBS17} and potentially even higher afterwards \citep{BCP18} - see, however, \citet{CDB13} and \citet{SBL15} - thereby shaping Earth's subsequent evolutionary history \citep{Holl06,LRP14,Knoll15}. If we choose the onset of the GOE as our critical step, we find that $\mathcal{P}_{2B} = 0.63$.\\

\noindent{{\bf (3B) Origin of Eukaryotes:}} We have already remarked previously as to why eukaryogenesis represented such an important step. The origin of eukaryotes, entailing the putative endosymbiosis of mitochondria (followed by the acquisition of plastids and other organelles), has gained near-universal acceptance as a seminal innovation from the standpoints of phagocytosis, gene expression, bioenergetics, organismal complexity and cellular evolution \citep{Mar81,PBB09,YWWK,Wag11,BSM15,MTM17,RMK17}, and is conventionally classified as a major evolutionary transition \citep{CS11}. The difference is that it constitutes the third step in our hypothesis, whereas it served as the second step in the original $5$-step model. The cumulative probability in this case is $\mathcal{P}_{3B} = 0.48$ since we have used the fact that eukaryogenesis occurred $1.8$ Ga based on our preceding discussion in step (2A).\\

\noindent{{\bf (4B) Origin of Complex Multicellularity:}} Our choice of (4B) is the same as (4A). This is primarily motivated by the fact that the origin of these organisms (especially plants and animals) have led to a radical transformation of Earth's biosphere. More specifically, Earth's energy balance, biomass productivity, biogeochemical cycles, ecological niches and macroevolutionary processes have been shaped by the emergence of complex multicellular organisms \citep{Lew00,OLF03,Butt07,PP09,Butt11,LUF15,Kno15}. Hence, in this case, we obtain the same cumulative probability, i.e. $\mathcal{P}_{4B} = \mathcal{P}_{4A} = 0.47$.\\
An alternative possibility is to consider the Neoproterozoic Oxygenation Event (NOE) as the critical step. The NOE is akin to the GOE since it also entailed a rise in the atmospheric oxygen (to near-modern levels), but its exact timing and causes are unclear. In particular, it remains ambiguous as to whether the NOE served as a cause or a consequence of the origin of animals \citep{OS12,LRP14}. The timing is also very variable, with evidence from selenium isotopes apparently not ruling out the onset of the NOE as early as $0.75$ Ga \citep{PS15} while iron- and iodine-based proxies seem to demonstrate significant oxygenation only as recently as $\sim 0.4$ Ga \citep{SW15,LRT18}. If we take the mean of these two quantities, the NOE would have taken place $0.55$ Ga and this estimate is roughly consistent with recent analyses that have yielded values of $\sim 0.5$-$0.6$ Ga \citep{CLV15,KN17}. If we assume the NOE to be a critical step instead, and use the value of $t = 3.95$ Gyr (i.e. $0.55$ Ga), we obtain $\mathcal{P}_{4B} = 0.57$.\\
At this stage, the following issue merits a clarification. We have argued earlier that the diversification of metazoans commenced at $0.8$ Ga \citep{CLBD17}, while the NOE has been assigned a timing of $0.55$ Ga. Hence, this raises the question as to how animal evolution took place in the presence of low oxygen levels. This discrepancy can be explained if the oxygen requirements for early animals (akin to modern demosponges) were sufficiently low \citep{SHK13,MWJ14,MC14,KS14}, or if the earliest metazoans were anaerobic altogether such as some species belonging to the phylum \emph{Loricifera} \citep{DD10,DG16}. Alternatively, factors other than oxygen - the most notable among them being, arguably, the availability of bioessential nutrients such as phosphorus \citep{RPG,Kno17,LS18,Ling18} - may have been responsible for regulating the advent of animals.\\

\noindent{{\bf (5B) Origin of Humans (Technological Intelligence):}} Our fifth step is essentially the same as that of the previous model (5A) on account of the following reasons. In addition to the distinctive ability to construct and utilize sophisticated tools (giving rise to complex information- and technology-driven networks), other attributes such as cumulative cultural transmission, social learning, mental time travel, large-scale social cooperation, mind reading, recursion and syntactical-grammatical language are also often cited as being unique to humans \citep{SC97,Dea98,PHP08,RB08,Tom08,Cor11,BRH11,WE12,Sud13,Tom14,HF14,HYB14,BC16,Hen16,Tom16,Lal17}.\footnote{On the other hand, it should also be appreciated that several ``human'' characteristics such as culture, intelligence, morality, foresight and consciousness have been, to varying degrees of controversy, associated with other species \citep{Gri01,RD05,WVS07,LaGa09,RF09,BP09,Row15,WR15,Roth15,DW16,DCD17}.} Lastly, humans have also caused major (perhaps irrevocable) large-scale shifts in the functioning of Earth's biosphere \citep{BMT11,BH12,EK13} to the extent that the Earth's current geological epoch, the Anthropocene, has been primarily shaped by us \citep{SG11,FS14,LM15,SBD15,WZ16}.\footnote{It is conceivable that the emergence of technological intelligence on other exoplanets may be accompanied by an equivalent Anthropocene epoch \citep{FKA17,FCAK}.} The cumulative probability for this step is given by $\mathcal{P}_{5B} = \mathcal{P}_{5A} = 0.37$.\\

\citet{ML10} introduced a parameter to estimate the goodness of fit:
\begin{equation}\label{DelDef}
    \delta = \frac{1}{n} \left[\sum_{r=1}^n \left(\mathcal{P}_{r,n} - 0.5\right)^2\right]^{1/2},
\end{equation}
and a lower value of $\delta$ corresponds to a better fit. If each of the cumulative probabilities $\left(\mathcal{P}_{r,n}\right)$ were close to either $0$ or $1$, it would mean that the events are clustered towards the beginning or the end, thereby constituting a poor fit. For the $5$-step model (1A-5A), we find $\delta_A = 0.068$. In contrast, if we use the $5$-step model (1B-5B), we find $\delta_B = 0.029$; even if use the GOE and the NOE in place of the steps (2B) and (4B) respectively, we find $\delta_B = 0.04$. Thus, we find that the second $5$-step model (1B-5B) is approximately twice more accurate than the first model (1A-5A).

\subsection{A six-step model}
\citet{Cart08} concluded that a 5- or 6-step model represented the best fit for the total number of critical steps on our planet. Apart from the two 5-step models delineated earlier, we note that another candidate is the ``energy expansions'' paradigm proposed by \citet{Jud17} that also involves $5$ steps in total. Here, we will outline a 6-step model based on the ``megatrajectories'' paradigm introduced by \citet{KB00} and assess whether it constitutes a good fit for the critical step framework.

\begin{itemize}
    \item \emph{From the Origin of Life to the Last Common Ancestor (LCA) of Extant Life:} As with the steps (1A) and (1B), we note that there is insufficient evidence to properly date the age when abiogenesis occurred and when the LCA lived. However, as we have argued in Sec. \ref{SSecFive}, the earliest definitive evidence for life appears to be around $3.7$ Ga. In this scenario, with $t_0 = 0.8$ Gyr and $t_H = 5.5$ Gyr, we use (\ref{CP}) to obtain $\mathcal{P}_{1,6} = 0.61$.
    \item \emph{The Metabolic Diversification of Bacteria and Archaea:} The first evidence for methanogens is arguably from hydrothermal precipitates dated $3.5$ Ga \citep{UYY06}, although molecular clock analyses lead to the even earlier date of at least $3.8$ Ga \citep{BFH04}. The earliest iron- and sulfate-reducing microbes also potentially appear in the fossil record at approximately the same time \citep{SBC01,UO08,WSB11,BSA12}. There is also some evidence suggesting that methanotrophy or the Wood-Ljungdahl pathway was operational at $3.4$ Ga \citep{FAS18,SKS18}. The record for nitrogen fixation implies that it was present by $3.2$ Ga \citep{SBG15}, or perhaps even earlier \citep{Stu16}. Thus, taken collectively there is considerable evidence indicating that metabolic diversification had occurred by $3.4$-$3.5$ Ga \citep{NCW13,Kno15,MJG17}. We will therefore adopt $t = 1.1$ Gyr (i.e. $3.4$ Ga), which results in $\mathcal{P}_{2,6} = 0.34$.
    \item \emph{Evolution  of  the  Eukaryotic Cell:} This megatrajectory is essentially the same as steps (2A) and (3B). Using the timing identified therein, we find $\mathcal{P}_{3,6} = 0.64$. 
    \item \emph{Multicellularity:} It is well-known that multicellularity has evolved repeatedly over Earth's history, and has been therefore characterized as a ``minor'' major transition \citep{GS07}. On the other hand, organisms that fall under the bracket of ``complex multicellularity'' belong to only six clades \citep{Kno11}. If the latter serves as the actual critical step, we have already discussed its timing in steps (4A) and (4B) and we end up with $\mathcal{P}_{4,6} = 0.23$.
    \item \emph{Invasion of the Land:} Although the first land-dwelling organisms appeared in the Precambrian \citep{WS15,DVKC}, the Paleozoic radiation of the land plants (embryophytes) facilitated a major ecological expansion. The earliest fossil evidence dates from the mid Ordovician \citep{Gen08}, although it is conceivable that land plants may have originated in the Cambrian \citep{KN17}. Consequently, the fossil record is in good agreement with molecular clock evidence that dates land plants to $0.45$-$0.55$ Ga \citep{STWB04,SBD10,MPC18}.\footnote{However, there are other molecular clock studies that favor a Proterozoic origin of land plants instead \citep{HGE01,CWD11,MHQ13}.} Thus, by choosing $t \approx 4$ Gyr, we find $\mathcal{P}_{5,6} = 0.48$.
    \item \emph{Intelligence and Technology:} This megatrajectory is essentially the same as steps (5A) and (5B). The corresponding cumulative probability is $\mathcal{P}_{6,6} = 0.30$.
\end{itemize}
By using (\ref{DelDef}), we compute the goodness of fit for this $6$-step model. We find that $\delta = 0.069$, which is virtually identical to $\delta_A$ (although lower than $\delta_B$ by a factor of about $2$). Hence, this demonstrates that the megatrajectories considered here are a fairly good fit insofar our model is concerned; the resultant value of $\delta$ is lower than the $5$- or $6$-step model analyzed in \citet{ML10}.

\subsection{The ramifications of an extended habitability interval}\label{SSecExt}
As noted in Sec. \ref{SSecNo}, recent theoretical studies indicate that the Earth may remain habitable (\emph{modulo} anthropogenic change) to $2$ Gyr in the future. With this revised estimate, the value of $t_H$ now becomes $6.5$ Gyr. As before, let us assume that humans represent the $n$-th step. By calculating the value of $n$ using (\ref{MeanT}), we find $n \approx 2.25$. Hence, this estimate suggests that a 2-step model (or possibly a 3-step one) has the greatest likelihood of being valid; \citet{Cart08} also reached a similar conclusion.  

We are confronted with the question as to what was the first critical step. The spacing between the critical steps must be approximately $2.2$ Gyr as seen from (\ref{Space}). Since the advent of humans at $4.5$ Gyr (i.e. $0$ Ga) constitutes the second step, the timing of the first critical step must have been approximately $2.2$ Ga. As noted in Sec. \ref{SSecFive}, the timing of the GOE (between $2.1$ to $2.4$ Ga) falls within this range. The GOE had profound consequences for Earth's subsequent evolutionary history, and therefore represents a strong contender for the first critical step. Other notable candidates that lie approximately within the same time frame include the evolution of (i) oxygenic photosynthesis and (ii) eukaryotes. In the $2$-step model, the origin of life (abiogenesis) is not likely to have been a critical step; in this regard, the $2$-step model is akin to the original $1$-step model proposed by \citet{Cart83}.

However, we can ask ourselves the following question: if the origin of life was a critical step, how many steps were there in total? If we assume that abiogenesis was the first step and that the mean time for this step was equal to the abiogenesis timescale of $0.8$ Gyr, from (\ref{MeanT}) we find $n \approx 7.1$. In contrast, if we had assumed that $t_H = 5.5$ Gyr and repeat the calculation, we arrive at $n \approx 5.9$. This leads us to the following conclusions:
\begin{itemize}
    \item For the case where habitability ends $1$ Gyr in the future, a $6$-step model would be favored, although the $5$-step model may also be plausible \citep{Cart08}. The choice of $n=5$ is consistent with previous analyses and the discussion in Sec. \ref{SSecNo}.
    \item When the habitability boundary extends to $2$ Gyr in the future, a $7$-step model would represent a good fit. Let us assume that $\bar{t}_{r,7} \approx 4.5$ Gyr, i.e. that humans are the $r$-th critical step. From (\ref{MeanT}), we obtain $r \approx 5.5$, implying that the evolution of humans could have been either the fifth or sixth critical step. In other words, there are still $1$ or $2$ critical steps ahead in the future, which we will discuss shortly hereafter.
\end{itemize}

As noted above, there is a possibility that humans are not the $n$-th critical step, but merely the $r$-th one (with $r \leq n$). In Secs. \ref{SSecNo} and \ref{SSecFive}, we have seen that there are compelling reasons to believe that humanity was the fifth critical step. Therefore, with $r = 5$ and assuming $\bar{t}_{5,n} \approx 4.5$ Gyr, we can estimate the value of $n$ using (\ref{MeanT}).
\begin{itemize}
    \item If $t_H = 5.5$ Gyr, and using the above values, we find $n = 5.1$. In other words, when Earth's habitability ends about $1$ Gyr in the future, the $5$-step model is relatively favored and the evolution of humans is the last critical step.
    \item Using the above parameters in conjunction with $t_H = 6.5$ Gyr leads us to $n \approx 6.2$. Hence, if the Earth becomes uninhabitable $2$ Gyr in the future, the $6$-step model seems the most likely. In this case, since humans are the fifth critical step, there is one critical step that is yet to occur.
\end{itemize}

Based on our discussion thus far, two broad inferences can be drawn. First, assuming that the habitability window ends $1$ Gyr in the future, the critical step model with $n=5$ is likely to be valid and humans represent the final critical step. In contrast, if the habitability window is extended to $2$ Gyr in the future, we suggest that the $6$-step model could be the best fit and that the rise of humans represents the fifth critical step. In other words, there is still one step in the future which is unaccounted for. Naturally, it is not possible to identify this critical step prior to its occurrence. 

One possibility is the emergence of superintelligence \citep{Bos14}, especially in light of recent advancements (and concerns) in Artificial Intelligence (AI) - the paradigm shifts presumably necessary for the genesis of ``human-like'' AI have been discussed in detail by \citet{LUTG}. However, the major issue from the standpoint of the critical step model is that the timescale between the first appearance of \emph{H. sapiens} and the dawn of AI superintelligence is currently predicted to be very low, i.e. on the order of $10^5$ to $10^6$ yrs, compared to the characteristic separation between successive critical steps ($\sim 10^9$ yrs); consequently, it remains unclear as to whether superintelligence can be regarded as a genuine critical step. This discrepancy might be resolved if the origin of superintelligence entails a much longer time than currently anticipated. 

An underlying assumption pertaining to the above discussion is that we have automatically presupposed that the biological ``complexity'' \citep{Carr01,MSB10} increases monotonically with time. The pitfalls of subscribing to implicit teleological arguments, certain theories of orthogenesis, and the ``March of Progress'' are many and varied \citep{Simp67,Ruse96,Gou96,Gould02},\footnote{Yet, many of the critical step models discussed in the literature take it for granted that the evolution of humans constitutes the last critical step regardless of the duration of the habitable period of the Earth.} and therefore it does not automatically follow that the sixth step alluded to earlier will lead to species of greater complexity. Hence, it does not seem implausible that the contrary could occur, especially if the environmental conditions $\sim 1$ Gyr in the future are less clement than today.

\section{Critical steps on exoplanets}\label{SSecExo}
We will now study some of the salient features of multi-step models on exoplanets, and discuss the resulting implications. A similar topic was studied recently (using the Bayesian framework) by \citet{Wal17} recently, but we incorporate additional constraints on habitability imposed by stellar physics in our treatment.

\subsection{The Habitable Zone of Earth-Analogs}

\begin{figure}
\includegraphics[width=7.5cm]{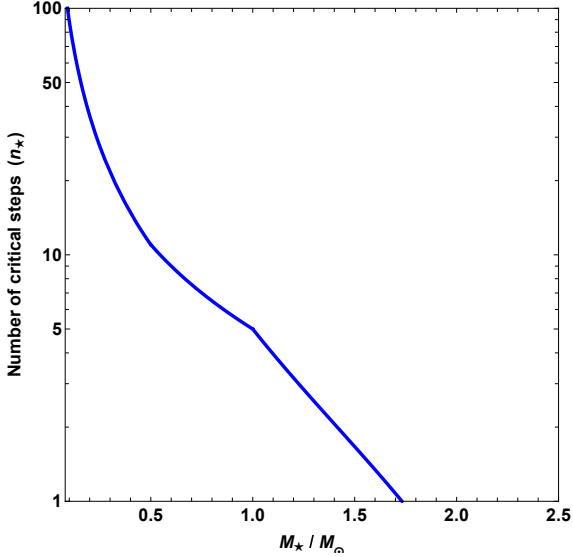} \\
\caption{Number of critical steps $n_\star$ as a function of the stellar mass $M_\star$ based on (\ref{nspec}).}
\label{FigMSpec}
\end{figure}

The habitable zone (HZ) is defined as the region around the host star where liquid water can exist on the planet's surface. The HZ is dependent on both planetary and stellar parameters, and evolves dynamically over time \citep{KWR93,KC03,KRK,KRS14}. The HZ is typically computed for ``Earth-analogs'', i.e. planets whose basic physical, chemical and geological parameters are similar to that of Earth. In our subsequent discussion, we will implicitly deal with Earth-analogs in the HZ of their host stars.\footnote{Thus, we shall not focus on habitable worlds outside the HZ, which are expected to be much more commonplace compared to those within the HZ \citep{LL17}.}

Clearly, the upper bound on the habitability of a planet is the stellar lifetime. However, the maximum duration that the planet remains habitable is less than the stellar lifetime for a simple reason: the stellar luminosity increases over time, and the planet will eventually enter a runaway greenhouse phase and become uninhabitable (like Venus). Thus, the duration of habitability is essentially specified by the temporal extent of the continuously habitable zone (denoted here by CHZ). By using the knowledge about the inner and outer boundaries of the HZ in conjunction with stellar evolution models, it is feasible to estimate the total duration of time $\left(t_{HZ}\right)$ that an Earth-analog will remain inside the HZ as a function of the stellar mass $M_\star$. This effort was undertaken by \citet{RCOW}, and by making use of Fig. 11 and Table 5 in that paper, we introduce the scaling:
\begin{eqnarray}\label{tHZ}
&& t_{HZ} \sim 0.55\,t_\odot\,\left(\frac{M_\star}{M_\odot}\right)^{-2} \quad \quad M_\star > M_\odot, \nonumber \\
&& t_{HZ} \sim 0.55\,t_\odot\,\left(\frac{M_\star}{M_\odot}\right)^{-1} \quad \quad\, 0.5 M_\odot < M_\star < M_\odot, \nonumber \\
&& t_{HZ} \sim 0.46\,t_\odot\,\left(\frac{M_\star}{M_\odot}\right)^{-1.25} \quad M_\star < 0.5 M_\odot,
\end{eqnarray}
where $t_\odot \sim 10$ Gyr and $M_\odot$ is the solar mass. Our choice of normalization constant differs from \citet{RCOW} since we have adopted the more conservative habitability duration of $5.5$ Gyr for the Earth-Sun system. By inspecting (\ref{tHZ}), it is apparent that low-mass stars are characterized by CHZs that last for a longer duration of time, which is along expected lines since they have longer main-sequence lifetimes \citep{AL97,LBS16}.

Now, let us consider the highly simplified model wherein we suppose that the timescale for abiogenesis is the same on all exoplanets, and that the duration of habitability is given by $t_{HZ}$. Since abiogenesis is taken to be the first critical step, from (\ref{MeanT}) with $r=1$ we find
\begin{equation}\label{nspec}
    n_\star = \frac{t_{HZ}}{t_\oplus}\left(n_\oplus + 1\right) - 1,
\end{equation}
where $t_\oplus = 5.5$ Gyr is the habitability duration of the Earth and $n_\oplus$ is the number of critical steps on Earth, while $n_\star$ represents the corresponding number of steps for Earth-analog orbiting a star of mass $M_\star$. We will henceforth use $n_\oplus = 5$ as this value has been advocated by several authors. Moreover, as we have seen from Sec. \ref{SecET}, there are reasons to believe that $n_\oplus = 5$ constitutes a fairly good fit. Thus, from (\ref{tHZ}), $n_\star$ can be estimated as a function of $M_\star$, and this plot is shown in Fig. \ref{FigMSpec}. The value of $n_\star$ decreases when $M_\star$ is increased, and shortly after $1.7\,M_\odot$ the value of $n_\star$ drops below unity.

\subsection{Constraints on habitability imposed by stellar physics}
The preceding analysis implicitly assumed that the only timescale for habitability was $t_{HZ}$. In reality, there are a number of factors governed by stellar physics that influence habitability \citep{LiLo}. In particular, there has been a growing appreciation of the role of space weather in governing habitability, i.e. for e.g. the role of stellar flares \citep{VKP17,OMJ17,MWW18}, coronal mass ejections \citep{KRL07,KOK16,DHL17}, stellar energetic particles \citep{SWM10,MLin17,LDF18,HTC18} and stellar winds \citep{VJM13,GDC16,GDC17,AG17,DLMC,DJL18,LiLo17} to name a few.

Since the presence of an atmosphere is necessary for maintaining liquid water on the surface of a planet, its complete depletion would lead to the termination of habitability insofar surficial life is concerned. For Earth-analogs that are closer to their low-mass host stars, they are subjected to intense stellar winds that can deplete their atmospheres over short timescales. The significance and magnitude of stellar wind erosion has been thoroughly documented in our Solar system \citep{BBMNS,Jak17}. The approximate timescale $t_{SW}$ associated with total atmospheric escape due to stellar wind erosion \citep{Man17} is
\begin{equation} \label{tSW}
t_{SW} \sim  100 t_\odot\, \left(\frac{M_\star}{M_\odot}\right)^{4.76},
\end{equation}
for an Earth-analog assuming that the star's rotation rate is similar to the Sun. The analytic model displays consistency with the trends and values discerned from numerical simulations \citep{DJL18,DLM18}. A few points should be noted regarding the above formula. First, incorporating the effects of stellar flares and coronal mass ejections \citep{DCY13,Cran17,OLH17,PG17,LDP18} is expected to decrease this timescale, perhaps by more than one order of magnitude. Second, this formula was derived under the assumption of a constant atmospheric escape rate. In reality, the escape rate may decrease by two orders of magnitude as the age of the star increases \citep{DLM18}. As a result, the higher escape rates in the past would further decrease the value of $t_{SW}$. Lastly, we observe that (\ref{tSW}) is applicable to unmagnetized planets. In contrast, if one considers strongly magnetized planets, the atmospheric escape rate could decrease by a factor of $\lesssim 10$ \citep{DLMC}, but the converse is also possible \citep{BT18,SST18}. In general, the escape rate is anticipated to be a non-monotonic function of the magnetic field \citep{GMN18,LiLo}.

Thus, as per the preceding discussion, the actual duration of habitability should be defined as $t_H = \mathrm{min}\{t_{HZ},t_{SW}\}$. In other words, if $t_{SW} < t_{HZ}$, the planet loses its atmosphere before it exits the CHZ and vice-versa. From (\ref{tHZ}) and (\ref{tSW}), we find
\begin{eqnarray}\label{tHab}
&& t_H \sim 0.55\,t_\odot\,\left(\frac{M_\star}{M_\odot}\right)^{-2} \quad \quad M_\star > M_\odot, \nonumber \\
&& t_H \sim 0.55\,t_\odot\,\left(\frac{M_\star}{M_\odot}\right)^{-1} \quad \quad\, 0.5 M_\odot < M_\star < M_\odot, \nonumber \\
&& t_H \sim 0.46\,t_\odot\,\left(\frac{M_\star}{M_\odot}\right)^{-1.25} \quad 0.41 M_\odot < M_\star < 0.5 M_\odot, \nonumber \\
&& t_H \sim 100 t_\odot\, \left(\frac{M_\star}{M_\odot}\right)^{4.76} \quad \quad  M_\star < 0.41 M_\odot.
\end{eqnarray}

\begin{figure}
\includegraphics[width=7.5cm]{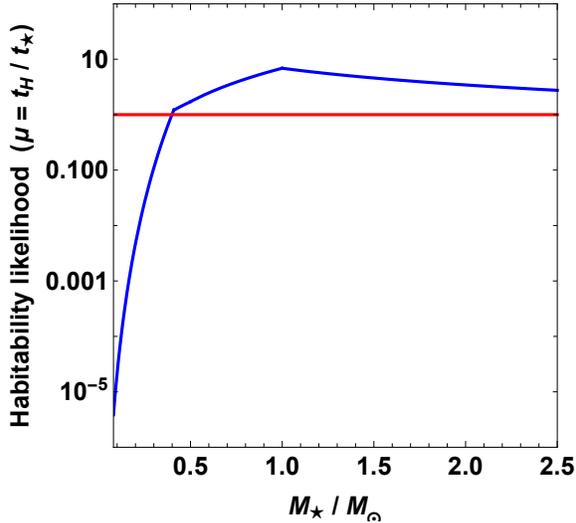} \\
\caption{The likelihood of a planet to host life $\mu$ (i.e. the ratio of its habitability duration to the abiogenesis timescale) as a function of the stellar mass $M_\star$.}
\label{FigHabLike}
\end{figure}

Our next stellar constraint stems from the availability of biologically active ultraviolet (UV) radiation. The importance of UV radiation partly stems from the fact that it constitutes the most dominant energy source for prebiotic synthesis on Earth \citep{CS92,DW10}. Although theories of abiogenesis are many and varied \citep{RBD14}, there is a strong case that can be made for UV radiation as the driver of prebiotic chemistry \citep{SK71,OML90,Pas12,McCo13,RV16,Suth17}, especially with regards to the RNA world \citep{Gil86,Joy02,Org04,NKB13,HL15,WAH17}, on account of the following reasons:
\begin{itemize}
    \item Laboratory experiments have shown that UV light provides a selective advantage to RNA-like molecules due to the presence of nitrogenous bases \citep{CCHK,GIM10,SSG16}, and may therefore play an important role in facilitating their oligomerization \citep{MCG03,DCG12}. 
    \item The observed tendency in myriad origin-of-life experiments to form complex organic mixtures incapable of Darwinian evolution is referred to as the ``asphalt problem''. Recent experiments have shown that this issue might be bypassed in suitable geological environments (e.g. intermountain valleys), and that UV radiation can potentially facilitate the synthesis of nucleosides, nucleotides, and perhaps RNA \citep{BKC12}.
    \item The synthesis of important biomolecules without excessive human intervention and under conditions that are presumed to resemble Hadean-Archean environments has proven to be challenging. However, there have been several breakthroughs in recent times that are reliant on UV light \citep{Suth16,IP17}. More specifically, the biologically relevant compounds produced include: (i)  pyrimidine ribonucleotides and $\beta$-ribonucleosides \citep{PGS09,XT17}, (ii) building blocks of sugars, such as glycolaldehyde and glyceraldehyde \citep{RS12,RS13,TFM18,XRR18}, (iii) precursors of nucleic acids, amino acids, lipids and carbohydrates \citep{BBG10,Pat15,RBL18}, and (iv) iron-sulfur clusters \citep{BVS17}.
    \item RNA nucleotides have been shown to be stable when radiated by UV photons, and this has been argued to be evidence that they could have originated in the high-UV environments of Hadean-Archean Earth \citep{SM09,GIM10,RT13,Beck16,RS16}. 
\end{itemize}
Another major theory for the origin of life posits that it occurred in submarine hydrothermal vents \citep{BH85,MB08}. This theory does have many advantages of its own \citep{MCS07,RBB14,SHW16}, and recent evidence suggesting that the LCA was thermophilic in nature is consistent with hydrothermal vents being the sites of abiogenesis \citep{ANY13,Weiss16}. However, it cannot be said at this stage that the LCA was definitively a thermophile, since other studies point to a mesophilic origin \citep{ML95,BL02,CF17}. A recent study by \citet{DD17} assessed seven factors ostensibly necessary for life's origination, and concluded that submarine hydrothermal vents may potentially face difficulties in fulfilling all of these criteria. 

It should be noted that UV radiation is expected to have other biological ramifications as well, both positive and negative. One of the downsides associated with high doses of UV radiation is that it can inhibit photosynthesis and cause damage to vital biomolecules (e.g., DNA) on Earth \citep{TS94,CSD05}. Yet, there are potential benefits stemming from UV radiation due to its proficiency in stimulating mutagenesis and thereby serving as a selection agent \citep{Sag73}. In particular, it is conceivable that UV radiation could have facilitated the emergence of evolutionary innovations such as sexual reproduction \citep{Roth99} and enhanced the rates of molecular evolution and speciation \citep{EG05}. It is therefore important to appreciate the possibility that there may be a number of advantages arising from UV radiation.

Hence, in our subsequent discussion, we will posit that the origin of life on Earth-analogs was driven by UV radiation. In this scenario, the rate of prebiotic chemical reactions is assumed to be constrained by the available bioactive UV flux at the surface \citep{BLM07,RWS17}. The latter can be estimated solely as a function of $M_\star$, thereby leading us to the abiogenesis timescale $t_\star$ \citep{Manasvi}:
\begin{eqnarray}\label{tOOL}
&& t_\star \sim t_0\,\left(\frac{M_\star}{M_\odot}\right)^{-3} \quad \quad M_\star \lesssim M_\odot, \nonumber \\
&& t_\star \sim t_0\,\left(\frac{M_\star}{M_\odot}\right)^{-1} \quad \quad M_\star \gtrsim M_\odot,
\end{eqnarray}
where $t_0 = 0.08 t_\odot = 0.8$ Gyr. Next, we consider the ratio $\mu \equiv t_H/t_\star$ because of its significance. If $\mu < 1$, then the duration of habitability is lower than the timescale for abiogenesis, thus implying that such Earth-analogs are not likely to host life. We have plotted $\mu$ as a function of $M_\star$ in Fig. \ref{FigHabLike}. A couple of conclusions can be drawn from this figure. For $M_\star \lesssim 0.4 M_\odot$, we find that $\mu < 1$ indicating that planets in the HZ of these stars have a lower chance of hosting life. Second, we find that the curve flattens out when $M_\star \gtrsim M_\odot$ but it does attain a slight peak at $M_\star = M_\odot$. Although this maximum is attained exactly at $M_\odot$ due to the ansatzen used in this paper, the peak of the curve has a high likelihood of being in the vicinity of $M_\star$, thereby suggesting that Sun-like stars may represent the most appropriate targets in the search for life \citep{LL18}.

In making use of (\ref{tOOL}), we have operated under the implicit assumption that stellar flares do not alter our results significantly. In actuality, stellar flares can deliver transient and elevated doses of UV radiation that are anticipated to have both positive and negative outcomes \citep{Dart11,MLin17,OMJ17,LiLo}. In particular, a recent analysis by \citet{RXT18} concluded that the background UV fluxes of stars with $T_\star < 4400$ K, with $T_\star$ denoting the effective stellar temperature, may not suffice for enabling UV-mediated prebiotic pathways to function efficiently. In some instances, however, stellar flares deliver sufficient UV photons to permit these reactions to occur. When the occurrence rate of flares ($\dot{N}_f$) exceeds the following threshold, UV-mediated prebiotic pathways could become functional.
\begin{equation}\label{FlaRatC}
   \dot{N}_f \gtrsim 3.36 \times 10^2\,\mathrm{day}^{-1}\, \left(\frac{E_f}{10^{34}\,\mathrm{erg}}\right)^{-1}\left(\frac{R_\star}{R_\odot}\right)^2 \left(\frac{T_\star}{T_\odot}\right)^4, 
\end{equation}
where $E_f$ is the flare energy and $R_\star$ is the stellar radius \citep{GZS19}. Based on the data from the \emph{TESS} mission, it has been estimated that $62$ stars out of a sample of $632$ flaring M-dwarfs fulfill the above criterion \citep{GZS19}. Thus, it might be reasonable to assume that the majority of low-mass stars do not flare frequently enough to compensate for the paucity of UV photons.

\begin{figure}
\includegraphics[width=7.5cm]{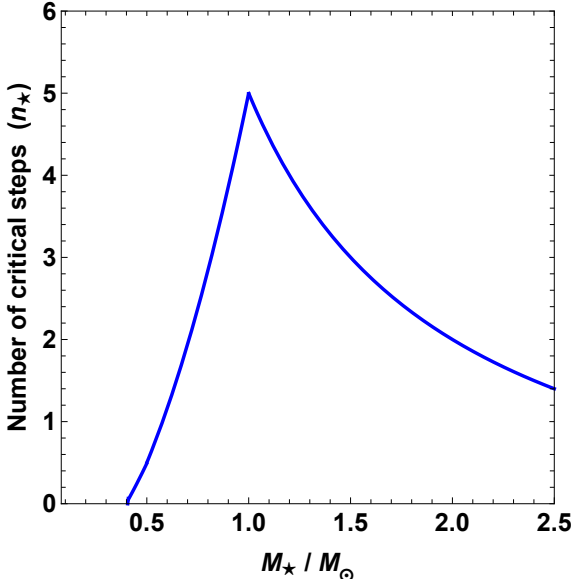} \\
\caption{Number of critical steps $n_\star$ as a function of the stellar mass $M_\star$ based on (\ref{nSup}).}
\label{FigHabSup}
\end{figure}

We turn our attention now to the $5$-step model introduced in Sec. \ref{SecET}. Since we have argued that abiogenesis was the first critical step, using (\ref{MeanT}) along with $t_H = 5.5$ Gyr for the Earth leads us to $\bar{t}_{1,5} \approx 0.92$ Gyr. Thus, we see that the timescale specified for the origin of life on Earth ($t_0 = 0.8$ Gyr) obeys $t_0 \approx \bar{t}_{1,5}$. As noted earlier, this is not surprising since the mean time taken for the $r$-th critical step in a viable model is comparable to its actual timescale \citep{Cart83}. If we assume that this condition is also valid on other potentially habitable planets, we have $\bar{t}_{1,n_\star} \approx t_\star$. Using this relation in conjunction with (\ref{MeanT}), (\ref{tHab}) and (\ref{tOOL}), the value of $n_\star$ is found to be
\begin{equation}\label{nSup}
    n_\star = \left(\frac{t_H}{t_\oplus}\right)\left(\frac{t_\star}{t_0}\right)^{-1}\left(n_\oplus + 1\right) - 1.
\end{equation}
Fig. \ref{FigHabSup} depicts the dependence of $n_\star$ on the stellar mass. From this plot, we see that $n > 0$ occurs only for $M \gtrsim 0.4 M_\odot$ and this result is in agreement with Fig. \ref{FigHabLike}, since planets orbiting such stars have a habitability duration that is shorter than the abiogenesis timescale. Second, we observe a clear peak at $M_\star = M_\odot$, and this behavior is also observed in some of the subsequent figures. This result is consistent with our earlier discussion: although the peak arises due to the scaling relations employed herein, there is still a strong possibility that the maximum number of critical steps occur when $M_\star \approx M_\odot$. It lends further credibility to the notion that G-type stars are the optimal targets in the search for life-bearing planets. Our results are qualitatively consistent with the Bayesian analysis by \citet{Wal17}, who concluded that: (i) the likelihood of life around M-dwarfs must be selectively suppressed, (ii) G-type stars are the most suitable targets for SETI (Search for Extraterrestrial Intelligence) observations, and (iii) the number of critical steps leading to intelligence is not likely to exceed five.

An important point to recognize here is that although the value of $n \approx 5$ occurs in the vicinity of $M_\star \approx M_\odot$, this does not altogether preclude stars outside this range from hosting planets with technologically sophisticated species. This is because the total number of critical steps leading to the emergence of life and intelligence on other planets is unknown, and there are no reasons to suppose the total number of critical steps will always be the same.\footnote{In light of the undoubted evolutionary and ecological significance of the breakthroughs discussed in Sec. \ref{SecET}, it may be tempting to conclude that they are sufficiently general, and argue that the convergent evolution of humanoids is ``inevitable''  if all these transitions are successful \citep{Mor03}. However, in spite of the impressive and rapidly increasing list of convergent mechanisms and organs \citep{McG11}, this standpoint appears to be overly anthropocentric.} On the other hand, once the number of critical steps drops below unity, it becomes rather unlikely that such stars (with $M_\star < 0.5\,M_\odot$) would have planets where intelligence can arise.

\begin{figure}
\includegraphics[width=7.5cm]{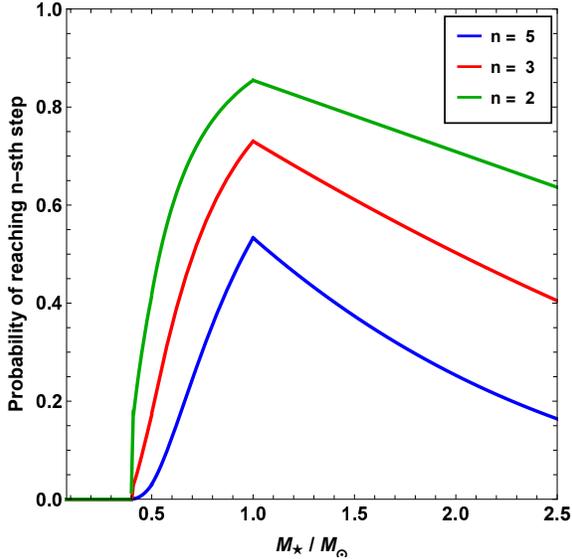} \\
\caption{The probability of attaining the $n$-step as a function of the stellar mass $M_\star$ based on (\ref{Pocc}).}
\label{FigProb}
\end{figure}

With these caveats in mind, we will, nevertheless, hypothesize that the critical steps discussed in Sec. \ref{SecET} (for Earth) are sufficiently general, and therefore applicable to other exoplanets. As we have seen, there are two constraints that were employed in our analysis: (I) the $n$ steps must occur in the interval $\left[0,t_H\right]$, and (II) the first step ($r=1$), namely abiogenesis, must occur at $t_0$. Hence, it follows that the remaining $n-1$ steps must unfold in remaining time interval. We introduce the expression for the PDF of the $\left(n-1\right)$-th step (in a sequence of $n-1$ steps) in the time $t'-t_0$:
\begin{equation}
    P_{n-1,n-1}\left(t'-t_0\right) = \mathcal{C} \left(t'-t_0\right)^{n-2},
\end{equation}
with $\mathcal{C}$ being a constant, based on Sec. \ref{SecMeth}. By integrating this PDF over the interval $\left[t_0,t_H\right]$, we obtain the probability, denoted by $\mathcal{P}_n\left(M_\star\right)$, for all $n$ steps to occur (since we have already imposed the constraint that the first step is attained at $t_0$). The constant of proportionality $\mathcal{C}$ is calculated by demanding that $\mathcal{P}_n\left(M_\star\right) = 1$ when $t_0 = 0$ because of criterion (I). This yields
\begin{equation}\label{Pocc}
    \mathcal{P}_n\left(M_\star\right) = \left(1 - \frac{t_0}{t_H}\right)^{n-1},
\end{equation}
and the same formula can be obtained from (\ref{CP}) with $r \rightarrow n-1$, $n \rightarrow n-1$ and $t \rightarrow t_H - t_0$; see also \citet{BT85}. Note that this formula is valid only when $t_0 < t_H$, which automatically excludes stars with $M_\star \lesssim 0.4 M_\odot$. There are two important scenarios worth considering from the standpoint of detecting the fingerprints of life:
\begin{itemize}
    \item {\bf The probability of technological intelligence:} This requires $n=5$ based on the above assumptions. In this case, it will be theoretically possible to detect signs of intelligent life by searching for technosignatures because they are more distinctive. 
    \item {\bf The probability of \emph{detectable} primitive life:} From the standpoint of microbial life, most of the well-known biosignatures like oxygen and ozone are not detectable until they have attained a certain level \citep{Mead17,KT18}. Hence, although Earth had life throughout most of its history, the low concentrations of oxygen and ozone until the GOE would have led to a ``false negative'' \citep{RO17}. Based on our discussion in Sec. \ref{SecET}, the evolution of oxygenic photosynthesis and the GOE correspond to $n=2$ (or $n=3$).
\end{itemize}
We have plotted (\ref{Pocc}) as a function of the stellar mass in Fig. \ref{FigProb}. It is seen that the peak is at $M_\star = M_\odot$, and that the curves rise sharply at $M_\star \approx 0.5 M_\odot$. The figure indicates that an Earth-analog around a G-type star would have the highest probability of achieving the critical steps necessary for detectable primitive or intelligent life. 

We must however point out an important caveat here. For $M_\star = M_\odot$ and $n = 5$, we obtain the probability $\mathcal{P}_n\left(M_\odot\right) \approx 0.5$ upon making use of (\ref{Pocc}). In other words, this result should imply that the likelihood of attaining the fifth critical step (intelligence), even with the constraints (I) and (II), is about $50\%$. Naturally, this value appears to be very high, but it must recognized that we have merely calculated the \emph{mathematical} probability. In reality, there will be a vast number of other criteria - for example, the presence of oceans (and continents), sufficiently high concentrations of bioessential elements, the existence of plate tectonics, the maintenance of a stable climate over Gyr timescales \citep{WaB00,Lam09,Kas10,JD10,MIG13,Ste16,Cock16,Ling18,ManLo18} - that must be simultaneously satisfied in order for each critical step to occur.\footnote{Although the total number of necessary and sufficient conditions that must have been fulfilled for all of the major transitions in Earth's history to occur was probably very high, especially if evolutionary contingency played a noteworthy role \citep{Sim64,Mon71,Mayr85}, we cannot say for certain whether every one of these evolutionary steps obeys the \emph{Anna Karenina} principle, i.e. the premise that the absence (or breakdown) of even a single factor is capable of dooming a particular process to failure \citep{Dia97}.} Hence, it is more instructive to view (\ref{Pocc}) as an upper bound, and use it to assess the relative chances of life-bearing planets existing around stars of differing masses.

\begin{figure}
\includegraphics[width=7.5cm]{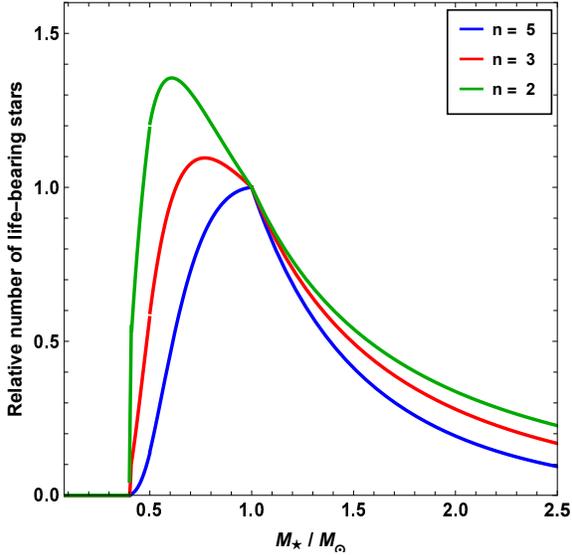} \\
\caption{The relative number of life-bearing stars $\left(\zeta_\star\right)$ which have completed $n$ critical steps as a function of the stellar mass $M_\star$.}
\label{RelNumb}
\end{figure}

Hitherto, we have only discussed the prospects for life on an Earth-analog around a given star. However, it should be recognized that the total number of stars also varies depending of their mass, i.e. low-mass stars are more numerous than high-mass ones \citep{BCM10}. Thus, we can calculate the relative number of stars $\zeta_\star = \mathcal{N}_\star/\mathcal{N}_\odot$ with detectable primitive or intelligent life, with the total number of stars $\mathcal{N}_\star$ defined as follows:
\begin{equation}\label{zetsta}
    \mathcal{N}_\star = \mathcal{P}_n\left(M_\star\right) \frac{d N_\star}{d\left(\ln M_\star\right)},
\end{equation}
where $dN_\star/d\left(\ln M_\star\right)$ represents the number of stars per logarithmic mass interval, and is calculated from the stellar initial mass function (IMF); here, we will use the IMF proposed by \citet{Kro01}. Note that $\mathcal{P}_n\left(M_\star\right)$ is given by (\ref{Pocc}) and can be viewed as a measure of the probability of planets with life (and have passed through the $n$ critical steps) per star.\footnote{We have not included an additional factor for the number of planets in the HZ of the host star because this quantity appears to be mostly independent of the stellar mass \citep{Kal17}.} 

We have plotted $\zeta_\star$ as a function of $M_\star$ in Fig. \ref{RelNumb}. Let us begin by observing that $\zeta_\star = 1$ when $M_\star = M_\odot$ by definition. For $n = 5$, we find that the peak occurs at $M_\star = M_\odot$, implying that solar-mass stars in our Galaxy are the most numerous in terms of planets with intelligent life. On the other hand, for $n = 2$, the peak is seen at $M_\star \approx 0.6 M_\odot$ (and at $M_\star \approx 0.77 M_\odot$ when $n = 3$) indicating that K-type stars are potentially the most numerous in terms of having primitive, but detectable, life. For the values of $n$ considered herein, we find that $\zeta_\star$ is almost constant in the range $0.5 M_\odot < M_\star < 1.5 M_\odot$ suggesting that these stars are the best targets in the search for life. 

The underlying reason for K-type stars being potentially more numerous in terms of hosting planets with detectable microbial biospheres stems from the fact that there are two distinct factors in (\ref{zetsta}). Hence, even though the probability per K-type star is comparatively lower than Sun-like stars - see Fig. \ref{FigProb} - the second factor (relative number of K-type stars) compensates for the first, namely the probability $\mathcal{P}_n\left(M_\star\right)$, in (\ref{zetsta}). On the other hand, when it comes to hosting planets with technological intelligence, the first factor in (\ref{zetsta}) dominates over the second, thereby ensuring that $\zeta_\star$ peaks for solar-type stars in Fig. \ref{RelNumb}.

\section{Conclusion}\label{SecConc}
We began by outlining a simple mathematical model predicated on the notion that evolution is effectively modeled as a series of independent ``hard'' steps. One of the primary objectives was to study the ramifications of this model for the timing and likelihood of primitive and intelligent life on Earth and Earth-like exoplanets around other stars. 

We began our analysis by focusing on the Earth and studying the total number of critical steps ($n$) that are likely on Earth based on the latest developments in geobiology. We found that the result depended on the time at which the Earth becomes uninhabitable in the future. For the more conservative estimate of $1$ Gyr, we found that $n=5$ probably represents the best fit, in agreement with previous studies. Unlike the standard $5$-step model \citep{Wat08,ML10} based on the classic paradigm of major evolutionary transitions \citep{JMS95,CS11,Sza15}, we proposed that the following five steps could have represented vital breakthroughs in the history of life on Earth: (i) abiogenesis, (ii) oxygenic photosynthesis, (iii) eukaryogenesis (endosymbiosis leading to the acquisition of mitochondria and subsequently plastids), (iv) complex multicellularity (e.g. animals and plants), and (v) genus \emph{Homo} (\emph{H. sapiens} in particular). On the other hand, if the Earth's habitability comes to an end $2$ Gyr in the future, we suggested that a $6$-step model might represent the best fit, wherein the emergence of humans constituted the fifth critical step with one major transition yet to occur in the future.

Subsequently, we applied this model to study the prospects for life on Earth-analogs orbiting stars of different masses. Our analysis took into account constraints based on: (i) the duration of the continuously habitable zone, (ii) atmospheric escape due to stellar wind erosion, and (iii) availability of bioactive UV flux to promote abiogenesis. We found that the timescale for abiogenesis is longer the duration of habitability for $M_\star < 0.4 M_\odot$, strongly suggesting that such stars are not likely to host life-bearing planets. The prospects for primitive or intelligent life are highest for a generic Earth-analog around a solar-mass star based on this analysis.

Next, we computed the total number of stars (relative to the solar value) that could give rise to detectable signatures of primitive and intelligent life. With regards to the former, we found that the number peaks in the range $0.6$-$0.8\,M_\odot$, implying that certain K- and G-type stars should potentially be accorded the highest priority in the hunt for biosignatures. Our analysis and conclusions are in agreement with previous studies of this subject \citep{Huang59,Dole,KWR93,HA14,TI15,CG16,Man17}. On the other hand, the total number of stars with intelligent life exhibited a peak near $M_\star \approx M_\odot$, thereby implying that Sun-like stars represent the best targets for SETI. This could also serve to explain why technological intelligence like our own finds itself in the vicinity of a solar-mass star, despite the fact that low-mass stars are more numerous and long-lived \citep{LBS16,HMK18}.

Naturally, there are a number of caveats that must be borne in mind with regards to the above conclusions. It is by no means clear as to whether evolution truly proceeds through a series of ``hard'' steps, and that the number and nature of these steps will be similar on other exoplanets. Our analysis has dealt solely with the stellar mass, although other stellar parameters (e.g. activity, rotation, metallicity) play an important role. Moreover, by focusing exclusively on Earth-analogs, we have not taken the intricate non-equilibrium biogeochemical factors that have shaped Earth's evolutionary history into consideration. Our discussion also ignored the possible transfer of life between stars. Such transfer may involve lithopanspermia \citep{Arr08,Bur04,Wick10}, directed panspermia \citep{CO73,Mau97} or interstellar travel and habitation by technologically advanced species \citep{SS66,Craw90,Lub16,Ling16,mlal18}. If such transfer events are common enough, which might be the case in some environments \citep{BMM12,DSR16,ManLi,CFL18,GLL18}, they could blur the quantitative conclusions of this paper because of diffusion processes and Galactic differential rotation \citep{NS81,ML16}. 

In spite of these limitations, it seems plausible that the critical step framework can be used to assess the relative merits of different models of the major evolutionary transitions on Earth. Furthermore, it also provides a useful formalism for gauging the relative likelihood of life on Earth-like planets orbiting different stars given the sparse data available at the current stage. Lastly, it offers testable predictions in the future, and, in principle, can therefore be falsified.

\section*{Acknowledgments}
ML is grateful to Andrew Knoll for the illuminating and thought-provoking conversations. This work was supported in part by the Breakthrough Prize Foundation for the Starshot Initiative, Harvard University's Faculty of Arts and Sciences, and the Institute for Theory and Computation (ITC) at Harvard University.


\end{document}